\documentclass[journal]{vgtc}                     
  \usepackage{booktabs}
  \usepackage{longtable}
  \usepackage{array}
  \usepackage{tabularx}
  \usepackage{fontawesome5}

  \usepackage{xcolor}
\newcommand\changed[1]{{\color{black}#1}}

\onlineid{0}

\vgtccategory{Research}

\title{AI-Centered Grand Challenges in Visual Analytics for Healthcare: Synthesizing the VAHC 2025 Community Experience}

\author{%
  \authororcid{J\"urgen Bernard}{0000-0001-8741-9709},
  \authororcid{David Gotz}{0000-0002-6424-7374},
  \authororcid{Robert S Laramee}{0000-0002-3874-6145},
  \authororcid{Silvia Miksch}{0000-0003-4427-5703},\\
  \authororcid{Gabriela Morgenshtern}{0000-0003-4762-8797},
  \authororcid{Renata G. Raidou}{0000-0003-2468-0664},
  and \authororcid{Alessio Arleo}{0000-0003-2008-3651}
}

\authorfooter{
  \item
  	J\"urgen Bernard and Gabriela Morgenshtern are with the University of Z\"urich.
  	E-mail: \{juergen.bernard|gabriela.morgenshtern\}@uzh.ch
  \item Silvia Miksch (E-mail: silvia.miksch@tuwien.ac.at) and Renata G. Raidou (E-mail: mail@renataraidou.com) are with the TU Wien.
  \item Robert S Laramee is with the University of Nottingham.
    E-mail: robert.laramee@nottingham.ac.uk
  \item David Gotz is with the University of North Carolina at Chapel Hill.
  	E-mail: gotz@unc.edu
  \item Alessio Arleo is with Eindhoven University of Technology.
    E-mail: a.arleo@tue.nl
}

\abstract{The intersection of AI, healthcare, and visualization is evolving rapidly, posing challenges that cut across disciplinary boundaries and resist easy resolution. 
The Visual Analytics in Healthcare workshop (VAHC), co-located every other year at the IEEE VIS conference and the AMIA (American Medical Informatics Association) annual conference, has served as a forum to connect the visualization and medical informatics community since 2010. 
In 2025, to celebrate the 16th edition, we used the workshop as an opportunity to consolidate the community's collective experience (and expertise) and identify Grand Challenges where the field should prioritize going forward. 
We combined thematic coding of the 15 accepted VAHC workshop papers with structured group discussions among more than 40 participants, organized around three major themes: ``Technical innovation vs.\ clinical reality", ``Human-centered and scalable VAHC", and ``From foundations to actionable insights", \changed{followed by post-workshop reflexive analysis.} 
Across all three groups, AI emerged as the most consistently recurring concern.
In this paper, we report our AI-centered insights from the VAHC 2025 group activity, contextualize them against the broader literature along five Grand Challenges themes, and distill them into five challenge clusters, each concluded with recommendations for future research directions that cross disciplinary boundaries: 
\textit{(1) trust and bias}, 
\textit{(2) data and infrastructure}, 
\textit{(3) explainability and communication}, 
\textit{(4) human-AI interaction}, 
and \textit{(5) model reliability and validation}. 
\looseness=-1
We share these challenges and their associated research directions as a starting point for discussion and collaboration across the healthcare, AI, and visualization communities.
\changed{All supplemental materials are available at \url{https://osf.io/p79uj}}.
}

\keywords{Visual Analytics, Visualization for Healthcare, Challenges and Opportunities, AI in Healthcare, Human-AI Interaction}

\graphicspath{{figs/}{figures/}{pictures/}{images/}{./}} 

\usepackage{tabu}                      
\usepackage{booktabs}                  
\usepackage{lipsum}                    
\usepackage{mwe}                       
\usepackage{ccicons}                   
\usepackage{xcolor}
\usepackage{graphicx}

\usepackage{mathptmx}                  

\begin{document}

\maketitle

\section{Introduction}
The rapid integration of AI into healthcare is reshaping how clinicians, patients, and researchers interact with health data. 
The visualization community is uniquely well-positioned to support this transition, yet faces the challenge of identifying where to focus its efforts.
The introduction of these technologies requires that researchers address unprecedented obstacles: those of scalability and data quality, resulting from the explosion in consumer-collected data, of those same consumers' sudden, unprecedented access to health data about themselves, of clinicians' skepticism towards AI recommendations in clinical decision-making and treatment planning, and of a lack of patient literacy, which complicates physician-patient relationships.

This is not the first time that the visualization community has challenged healthcare research. 
In 2015,  Caban and Gotz~\cite{cabanG15} recognized visualization and visual analytics (VA) as disciplines capable of realizing the potential behind the adoption of Electronic Health Records (EHR) on a large scale, rejecting paper records as the standard for recording and sharing clinical data.
This ``call to action" was written to provoke a reaction in both the visualization and medical informatics communities, pushing the two towards deeper collaboration on research, towards more open, accessible, and personalized care. 
The Visual Analytics in Healthcare (VAHC) workshop~\cite{vahc2025web} is a forum where the visualization and medical informatics communities can gather to discuss research intersecting visual data analysis and clinical informatics. 
Topics of interest to the workshop include applications of VA to clinical care, public health data analysis, AI for healthcare data visualization, and patient- and consumer-related methods. 
In 2024, a special issue of the \textit{Journal of American Medical Informatics Association} (JAMIA) was published on the topic of digital and personal health, following the results of VAHC 2024~\cite{jamiaEditorial2024}.

As with the introduction of EHR, the advent of AI methods in healthcare has sparked a radical paradigm shift, impacting various aspects of research and practice (see, e.g., ~\cite{reardon_rise_2019,singh_artificial_2023, shang_artificial_2024}). 
These include support in patient diagnosis and clinical decisions~\cite{shang_artificial_2024}, automated identification of cancerous tissues from X-rays~\cite{reardon_rise_2019}, drug research~\cite{singh_artificial_2023}, robot-assisted surgical procedures~\cite{shang_artificial_2024}. 
Within this "revolution", with opportunities and challenges on the horizon, VAHC 2025 set out to directly involve the community in a collaborative effort to identify the \textit{Grand Challenges} in VAHC, intended to cross boundaries between visualization/VA and medical informatics communities. 

In this paper, we report the AI-centered findings from VAHC 2025, co-located with IEEE VIS.
Our process is grounded in over a decade of VAHC workshop proceedings, and includes systematic thematic coding of the 15 accepted VAHC 2025 submissions, structured group discussions with more than 40 workshop participants, and post-workshop reflexive analysis. 
This process reveals five AI-centered challenge clusters (bias and trust, data and infrastructure, explainability and communication, human-AI interaction, and model reliability and validation) that recurred independently across three differently-framed discussion groups, suggesting they reflect community-wide priorities rather than organizer framing. 
Across all five, a consistent signal emerges: the most persistent AI challenges in healthcare VA are not technical, but rather calibration problems: between trust and evidence, between explainability and clinical utility, between automation and human agency. 
We offer these clusters and their associated research directions as a synthesis of current community concerns and a step toward a shared research agenda at the intersection of healthcare, AI, and visualization.

\section{Methods}
\label{sec:methods}
\textbf{Longitudinal Literature Grounding.}
We base our approach on over a decade of VAHC workshop proceedings, ensuring that VAHC 2025 grand challenges activities will be well-grounded.
Caban and Gotz~\cite{cabanG15} presented a structured survey of open problems in health VA, identifying integration with clinical workflows, scalability, evaluation, and data privacy as persistent gaps, many of which remain unsolved. 
A subsequent VAHC special issue~\cite{jamiaEditorial2024} updated this inventory around six clusters: explainable and trustworthy AI, user-centered design, health communication, EHR accessibility and privacy, equity and inclusion, and interdisciplinary collaboration.
External perspectives on evaluation rigor, encoding standards, and the gap between visualization research norms and clinical evidence requirements were additionally drawn from a panel on health visualization challenges at EuroVis 2025~\cite{euroVisPanel2025}. 
Together, these sources informed the controlled vocabulary used for submission coding and the three Grand Challenge themes.

\textbf{Pre-Workshop Submission Coding.}
We coded all 15 accepted VAHC 2025 papers (seven research papers, four posters and system demonstrations, and four Grand Challenge papers) against a controlled vocabulary of 20 thematic codes, 
derived inductively from the submissions and anchored in the literature above. 
\changed{Codes included evaluation methodology, Human-centered AI (HCAI), LLM applications, decision-making support, user task characterization, scalable data analysis, wearable data, genomics, reasoning under uncertainty, and visualization fundamentals, among others.} 
This coding provided a transparent, assumption-independent basis for identifying workshop themes and allocating submissions to discussion groups, reducing the risk that theme selection reflected organizer bias rather than the community's actual research priorities.
Coding was performed independently by one VAHC conference organizer (co-author), with all other organizers reviewing assignments; disagreements were resolved through group discussion until consensus was reached.

\textbf{Theme Derivation and Workshop Execution.}
Similarity-based clustering of the coded submissions yielded three themes. 
\begin{itemize}
    \item \textit{Theme 1} (Technical Innovation vs. Clinical Reality) clustered around LLM applications, HCAI, and clinical decision-making support, addressing the persistent gap between academic innovation and clinical adoption. 
The discussion prompt asked: \textit{How can we ensure that the tools we innovate are clinically viable, trusted, and adoptable by end users?} 
    \item \textit{Theme 2} (Human-Centered Scalable VAHC) addressed user task characterization, 
    wearable and genomic data, and personal-to-population scaling. The discussion prompt asked: \textit{How do we design visualization frameworks that serve multiple stakeholders while scaling from individual patient care to population health?} 
    \item \textit{Theme 3} (From Foundations to Actionable Insights) consolidated visualization fundamentals, evaluation methodology, and reasoning under uncertainty. The discussion prompt asked: \textit{What design principles and evaluation standards ensure healthcare visualization tools reliably guide users from exploratory analysis to actionable decisions?} 
\end{itemize}
Formulating each theme as an open prompt rather than a closed definition was a deliberate choice to encourage participants to surface unexpected directions rather than confirm pre-existing assumptions.

The activity portion of the workshop opened with four lightning talks presenting the accepted Grand Challenge papers ~\cite{wang2025vahc,ershova2025vahc,assor2025vahc_changing,vonwyl2025vahc}, each framing an open research problem and further operationalizing the tension between technical innovation and clinical adoption (see Table~\ref{tab:mapping}. 
Ultimately, these talks seeded the structured group discussions that followed, in which more than 40 participants engaged in 90-minute sessions per theme, using hand-written notes to identify challenges, sub-themes, and research directions, to lower the barrier to participation. 
Gotz, Bernard, and Miksch moderated the groups; designated note-takers Arleo, Morgenshtern, and Laramee recorded and summarized outcomes. 
\changed{Participants spontaneously joined the groups based on their personal interests in the health-related themes. 
The groups were diverse in healthcare-real-world experience, comprising academic participants at professor, postdoctoral researcher, and PhD student level. }

\textbf{Post-Workshop Analysis and AI Scope Decision.} Following the workshop, each moderating pair conducted reflexive thematic analysis of their group's notes, breaking down discussions into overarching topics. 
The full author team then collaboratively analyzed all three sets of discussion notes, iteratively identifying convergences across themes. 
Convergence across independently moderated groups strengthened confidence that the identified topics reflect community-wide priorities rather than the perspective of any single discussion group.
\changed{This process resulted in 155 statements being identified across the three themes, with 53, 62, and 40 per group, respectively.}
AI emerged as a recurring theme across all three groups, as well as across the other VAHC discussion formats (submissions and lightning talks). 
From the full set of VAHC workshop discussion notes from the group activity, we isolated AI-related statements (69/155) and applied a bottom-up inductive coding approach, deriving codes and sub-codes from the data and resulting in five overarching codes, each with three to four sub-categories \changed{(see Table~\ref{tbl:aiutterances}). 
A detailed breakdown of statements and coding is available as supplemental material.}
Consistent with reflexive thematic analysis practices, members of the research team acted as active interpreters in this process. 
The coding process involved discussions to reflect on and refine interpretation by the entire team.
For each resulting cluster, we examined its presence across workshop submissions, contextualized findings against the broader literature, and distilled opportunities and future directions to seed cross-disciplinary research agendas.

\changed{
\begin{table}[t!]
    \centering
    \caption{Distribution of AI statements per challenge cluster.}
    \vspace{-2mm}
    \label{tbl:aiutterances}
    \setlength{\tabcolsep}{6pt}
    \renewcommand{\arraystretch}{0.95}
    \begin{tabularx}{\columnwidth}{@{} c X r @{}}
        & \textbf{Challenge Cluster} & \textbf{Statements} \\
        \midrule
        \includegraphics[height=1.0em]{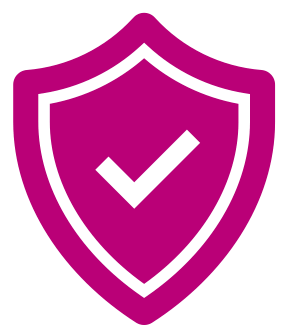} & Bias and Trust & 12 \\
        \includegraphics[height=1.0em]{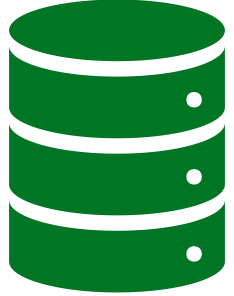} & Data and Infrastructure & 12 \\
        \includegraphics[height=1.0em]{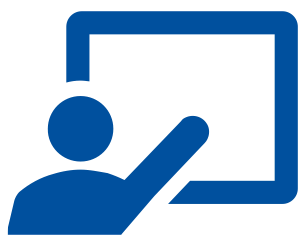} & Explainability and Communication & 12 \\
        \includegraphics[height=1.0em]{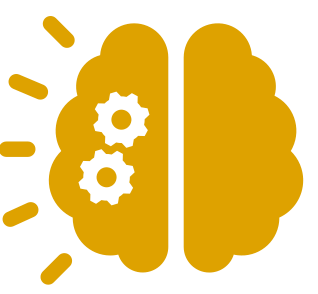} & Human-AI Interaction & 23 \\
        \includegraphics[height=1.0em]{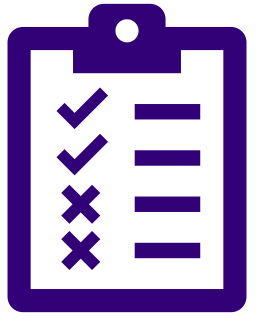} & Model Reliability and Validation & 10 \\
        \midrule
        & \textbf{Total} & \textbf{69} \\
        \bottomrule
    \end{tabularx}
    \vspace{-6pt}
\end{table}
}

\section{AI-Centered Grand Challenges in VAHC}
\label{sec:AIchallenges}
We present five challenge clusters that emerged from our analysis \href{https://docs.google.com/spreadsheets/d/1Onj7iHTkEscuBzHxWplloRb9OL9cvSSgdYUcQPti-Xg}{\faExternalLink*}.
Table~\ref{tab:mapping} \changed{maps VAHC 2025 workshop submissions to} the challenges. 
We discuss each challenge cluster through workshop findings, broader literature, and conclude with concrete opportunities and directions.

\paragraph{\includegraphics[height=1.1em]{figs/icons/trust.png} Bias and Trust}
The workshop discussion on bias and trust (12/69) highlights three core challenges for healthcare. 
First, multiple forms of \textit{bias} (5/12)---model, data, and human (e.g., confirmation bias)---raise concerns about reliability and decision quality, emphasizing the need for lifecycle-wide mitigation. 
This concern aligns with work documenting how biases compound across the AI development pipeline, from data acquisition through deployment, with downstream effects on clinical decision quality and healthcare disparities \cite{cross2024bias, chen2023algorithmic}.
Second, \textit{misinformation} (3/12) emerges both from public misconceptions about AI (e.g., equating all AI to LLMs) and from external sources, such as forums, exacerbated during crises such as COVID-19~\cite{arleoBKRSY25}. 
Third, \textit{trust calibration} (4/12) is critical: both overtrust (even among experts) and distrust must be managed, promoting informed skepticism. 
\changed{In visualization research, overtrust was empirically shown to increase with task difficulty, and transparency alone does not reliably recalibrate it~\cite{ha2024guidedbyai}, a finding with direct implications for time-pressured clinical settings.}
Workshop contributions illustrate these tensions in deployed settings. 
Nipu et al.~\cite{nipu2025vahc} report on the post-deployment evaluation of an interactive oncological risk estimator, documenting how clinician feedback on a multi-view AI-backed dashboard shapes revisions to risk communication and cohort-level stratification. 
Hosseini et al.~\cite{hosseini2025vahc} examine how LLMs adapt visualization rhetoric to population health personas, raising open questions about how persona-driven rhetorical choices may reinforce stakeholder priors and affect calibrated interpretation.
In the group activity, VAHC systems were seen as potential tools for communicating and, therefore, strengthening trust between patients, clinicians, and other healthcare providers, as well as a way to reduce or raise awareness about potential biases. 
\looseness = -2
Towards this goal, Sacha et al.'s conceptual framework~\cite{sacha_trust-VA_2025} proposes decomposing uncertainty propagation throughout the VA pipeline and process, providing a basis for explicitly communicating uncertainty at key stages and introducing operator- or system-level interventions that counteract perceptual and cognitive biases.
Such an approach, instantiated in healthcare contexts~\cite{musleh2025trustme}, can directly support calibrated trust in clinical VA systems.

\changed{
\begin{table}[ht]
\setlength{\tabcolsep}{4pt}
\renewcommand{\arraystretch}{1.15}
\centering
\caption{Author labels and paper titles mapped across the five identified AI-centered challenges: \includegraphics[height=1.1em]{figs/icons/trust.png} \textit{Bias and Trust}, \includegraphics[height=1.1em]{figs/icons/data.png} \textit{Data and Infrastructure}, \includegraphics[height=1.1em]{figs/icons/explainability.png} \textit{Explainability and Communication}, \includegraphics[height=1.1em]{figs/icons/humanAI.png} \textit{Human-AI Interaction}, \includegraphics[height=1.1em]{figs/icons/validation.png} \textit{Model Reliability and Validation}.}
\vspace{-2mm}
\label{tab:vahc-author-title-mapping}
\begin{tabularx}{\linewidth}{@{} >{\hsize=0.55\hsize}X >{\hsize=1.45\hsize}X @{}}
  \toprule
  \textbf{Author} & \textbf{Title} \\
  \midrule
  \footnotesize Nipu et al.\hfill\includegraphics[height=1.1em]{figs/icons/trust.png}\includegraphics[height=1.1em]{figs/icons/explainability.png}
    & \footnotesize Lessons from the Development and Deployment of an Interactive Oncological Risk Estimator~\cite{nipu2025vahc} \\
  \footnotesize Assor et al.\hfill\includegraphics[height=1.1em]{figs/icons/data.png}\includegraphics[height=1.1em]{figs/icons/humanAI.png}
    & \footnotesize Understanding Expert Exploration in EHR Visualization Tools: The ParcoursVis Use Case~\cite{assor2025vahc_understanding} \\
  \footnotesize Ershova et al.\hfill\includegraphics[height=1.1em]{figs/icons/data.png}
    & \footnotesize \textit{[Grand Challenge]:} Bridging Service Design, Visualizations, and Visual Analytics in Healthcare Digital Twins~\cite{ershova2025vahc} \\
  \footnotesize Rauscher et al.\hfill\includegraphics[height=1.1em]{figs/icons/data.png}
    & \footnotesize MotiVAtor: Analyzing Physical Activity Study Data in Lab and Life through Visual Analytics~\cite{rauscher2025vahc} \\
  \footnotesize Aupetit et al.\hfill\includegraphics[height=1.1em]{figs/icons/data.png}
    & \footnotesize InViTAG: a web application for AI-assisted exploration and grouping of health images and data~\cite{aupetit2025vahc} \\
  \footnotesize Balaka et al.\hfill\includegraphics[height=1.1em]{figs/icons/explainability.png}
    & \footnotesize The MoBa GWAS Explorer: Designing Approachable Visualizations of GWAS Data for a Mixed Audience~\cite{balaka2025vahc} \\
  \footnotesize Owen et al.\hfill\includegraphics[height=1.1em]{figs/icons/explainability.png}
    & \footnotesize Embedding Empathy into Visual Analytics: A Framework for Person-Centred Dementia Care~\cite{owen2025vahc} \\
  \footnotesize Assor et al.\hfill\includegraphics[height=1.1em]{figs/icons/data.png}\includegraphics[height=1.1em]{figs/icons/humanAI.png}
    & \footnotesize \textit{[Grand Challenge]:} Changing the Paradigm from Dynamic Queries to LLM-generated SQL Queries with Human Intervention~\cite{assor2025vahc_changing} \\
  \footnotesize Hosseini et al.\hfill\includegraphics[height=1.1em]{figs/icons/trust.png}\includegraphics[height=1.1em]{figs/icons/humanAI.png}
    & \footnotesize What Do LLMs Prioritise When Adapting Visualizations to User Personas?~\cite{hosseini2025vahc} \\
  \footnotesize Lee et al.\hfill\includegraphics[height=1.1em]{figs/icons/humanAI.png}
    & \footnotesize VAIR: Visual Analytics for Injury Risk Exploration in Sports~\cite{lee2025vahc} \\
  \footnotesize Kleinau et al.\hfill\includegraphics[height=1.1em]{figs/icons/humanAI.png}
    & \footnotesize A Dashboard for Visualizing Health Data of Children with Aortic Valve Disease~\cite{kleinau2025vahc} \\
  \footnotesize Pakianathan et al.\hfill\includegraphics[height=1.1em]{figs/icons/humanAI.png}
    & \footnotesize Exploring Human-AI Interaction with Patient-Generated Health Data Sensemaking for Cardiac Risk Reduction~\cite{pakianathan2025vahc} \\
  \footnotesize von Wyl et al.\hfill\includegraphics[height=1.1em]{figs/icons/trust.png}\includegraphics[height=1.1em]{figs/icons/validation.png}
    & \footnotesize \textit{[Grand Challenge]:} Mediating Research Cultures of Health Sciences and Visual Analytics~\cite{vonwyl2025vahc} \\
  \footnotesize Shell et al.\hfill\includegraphics[height=1.1em]{figs/icons/explainability.png}\includegraphics[height=1.1em]{figs/icons/validation.png}
    & \footnotesize Explaining Model Predictions When Users Aren't Sure: XAI for scRNA-seq Cell-Type Classification~\cite{rl2025vahc} \\
  \footnotesize Wang et al.\hfill\includegraphics[height=1.1em]{figs/icons/explainability.png}\includegraphics[height=1.1em]{figs/icons/validation.png}
    & \footnotesize \textit{[Grand Challenge]:} Visual Analytics for Causal Reasoning from Real-World Health Data~\cite{wang2025vahc} \\
  \bottomrule
\end{tabularx}
\label{tab:mapping}
\vspace{-5mm}
\end{table}
}

\noindent \textbf{\emph{Opportunities and Directions.}}
Across our sub-themes, biases and trust emerge as situated constructs shaped not only by data and model performance, but by how uncertainty, bias, and provenance are communicated in context. 
This points to the need for standardized frameworks for trust calibration that explicitly account for both overtrust and undertrust in deployed clinical AI systems, and that guide how these factors should be communicated and operationalized in clinical practice.

\paragraph{\includegraphics[height=1.1em]{figs/icons/data.png} Data and Infrastructure}
The workshop discussion topics on data and infrastructure (12/69) are equally prominent in AI-driven healthcare. 
Large-scale \textit{data collection} (3/12)---from patient journaling to continuous sensing---offers significant potential, but creates analytical bottlenecks and data quality problems, \changed{with most data remaining underutilized or not validated according to quality metrics or medical domain knowledge.} 
This highlights the need for effective pipelines that transform raw data into clinically meaningful insights.
Workshop contributions also underscore the opportunity to better leverage ubiquitous sensing for the benefit of patient care. 
Assor et al. \cite{assor2025vahc_understanding} present ParcoursVis, a progressive VA system operating on EHR-derived event sequences at the scale of millions of patients and billions of events, and report an insight-based evaluation protocol developed with hospital practitioners. 
Rauscher et al. \cite{rauscher2025vahc} address the analytical challenges of multigranular physical activity datasets spanning laboratory and real-life wearable measurements, contributing a linked-view application for hypothesis generation in interdisciplinary intervention studies.
At the same time, \textit{governance} concerns (5/12) emerge around data trustworthiness, privacy, and access, alongside calls for standards such as FAIR principles~\cite{wilkinson2016fair}. 
While shared \changed{data access across a care team can aid effective care, access restrictions in the healthcare system exist (in part) to protect patients, and care providers must uphold secure, appropriately restricted access.} 
Persistent gaps in data stewardship were found during the COVID-19 pandemic across clinical information systems \cite{queralt2022fair}, echoing the presence of this challenge externally from the VAHC research community.
Finally, \textit{integration} (4/12) remains a key challenge, requiring interoperability across heterogeneous sources and a clearer data infrastructure to enable seamless, reliable use in clinical and patient-centered contexts. 
In the group activity, VAHC systems were seen as enablers of integration, exploration, and interpretation of heterogeneous data streams, transforming large-scale raw data into actionable insights while supporting transparency, governance, and interoperability across clinician- and patient-centered systems.
Data provenance communication strategies in VA can address issues around accountability and liability present when clinicians begin to engage with patient-provided data, providing information on data sources (patient, clinical, sensor-based vs. qualitative). 
\looseness = -2
Systematic review evidence indicates that healthcare professionals and researchers generally view patient-generated health data from wearables and apps positively, while concerns regarding data reliability and integration of heterogeneous data sources into existing clinical workflows continue to constrain clinical adoption \cite{bruckner2025attitudes, wang2022ehr}.

\noindent \textbf{\emph{Opportunities and Directions.}}
Rather than data scarcity, the primary opportunity lies in addressing fragmentation and governance. 
Future work focuses on developing end-to-end, interoperable pipelines that not only integrate heterogeneous data sources but also make data quality, provenance, and uncertainty explicit throughout.
\changed{Specifically, provenance~\cite{xu2020provenance} uncertainty~\cite{ristovski2014uncertainty} visualization state-of-the-art reports offer a robust conceptual foundation for transparent provenance tracking that support accountability and actionable clinical decision-making.}

\paragraph{\includegraphics[height=1.1em]{figs/icons/explainability.png} Explainability and Communication}
Among the statements centered on explainability and communication (12/69), a key focus is effective \textit{risk communication} (3/12). 
This includes how to present uncertainty and establish standardized, evidence-based methods for conveying risk to patients. 
\textit{Visualization} (3/12) plays a critical role in making data provenance, quality, and insights understandable. 
At the same time, limitations of transparency and tensions between \textit{explainable AI (XAI)} research and clinical practice emerge (6/12), where explainability is often secondary to validated outcomes. 
Critiques that post-hoc explanations of black-box models provide limited epistemic guarantees for high-stakes clinical decisions motivate the use of inherently interpretable models where feasible \cite{rudin2019stop}.
For example, work such as Ghassemi et al.'s \cite{ghassemi2021falsehope} argues that current XAI methods are unlikely to deliver the trust, transparency, and bias-mitigation outcomes frequently attributed to them at the patient level, and \changed{recommends} rigorous internal and external validation as a more direct path to clinical utility.
\looseness-1
Broader reflections emphasize a shift toward human-centric \changed{(explanatory) AI~\cite{meske2025explainable}}, rather than model-centric notions of explainability, questioning what “explainable” means in context. 
This includes comparing AI to human decision-making and highlighting the need for context-sensitive, task-oriented explanations. 
In VAHC 2025, Balaka et al. \cite{balaka2025vahc} present the MoBa GWAS Explorer, a design study investigating onboarding strategies that render genome-wide association study data tractable for both expert and non-expert audiences drawn from a longitudinal cohort study. 
Owen et al. \cite{owen2025vahc} propose an empathy-centred visualization framework for person-centred dementia care, extending explainability from model reasoning to the communication of indirect end-user experience.
In our group activity, VAHC systems appear as answers to making AI decisions interpretable by clearly communicating risk, uncertainty, and data provenance through context-sensitive, task-oriented visual interfaces. 
A key direction emerging here is the need for systematic, context-aware approaches to communicating data, decisions, and interventions for interpretability. During COVID-19, this need became particularly visible. 
For instance, Baumgartl et al.~\cite{baumgartl_COVID_2025} systematically characterize user groups, task structures, modeling approaches, and data modalities, and analyze the visual encodings employed to communicate data, decisions, and interventions in support of explainability. 
\changed{Xu et al.~\cite{xu_application_2018} introduce a study concerning the application of data provenance in the design of a healthcare analytics software, mindful of the complex human cognitive activities involved in the process and the challenges of presenting such complex information even to expert users.}

\noindent \textbf{\emph{Opportunities and Directions.}}
Closing the gap between technical explainability and clinical meaningfulness, where outcomes and usability consistently precede formal model interpretability in practitioners' priorities~\cite{buhler2025aiintheloop}, is critical. 
This calls for visualization approaches that communicate uncertainty, provenance, and bias in context, evaluated for their measurable impact on decision quality and user confidence.

\paragraph{\includegraphics[height=1.1em]{figs/icons/humanAI.png} Human-AI Interaction}
One third of the statements (23/69) highlight how Human-AI interaction reshapes access, roles, and workflows in healthcare. 
AI systems, like LLMs, offer \textit{accessibility} (6/23) to communication, enabling openness and supporting patient education, though challenges remain around appropriate abstraction for diverse users. 
At the same time, \textit{roles} (4/23) are shifting. 
Clinicians may act as designers or supervisors, raising concerns about de-skilling and the need for clear human oversight and feedback loops. 
In the literature, a proposed way to mitigate this risk and ensure that agency and AI accountability remain with humans is to use visualizations as the integration layer, interfacing AI with clinicians and leveraging an "AI-in-the-loop" paradigm \cite{buhler2025aiintheloop}.
In terms of \textit{use} (5/23), effective interaction depends on prompt design, multimodal interfaces, and ensuring AI augments rather than replaces expertise. 
Finally, \textit{workflows} (8/23) are evolving toward collaborative, distributed decision-making, where AI supports both clinicians and patients across care journeys, including onboarding and guidance~\cite{stoiber_onboarding-guidance-2022}, and community-driven tools. 
However, \changed{evaluations remain limited in generalizability, motivating future work that explores shared design frameworks} for clinical AI workflows \cite{lai2023science}.
Among the VAHC papers, Lee et al. \cite{lee2025vahc} exemplify this distributed configuration outside the clinic with VAIR, a VA system that couples pose estimation and biomechanical simulation with synchronized views of joint-level risk indicators, supporting expert reasoning for both retrospective injury analysis and proactive intervention planning in sports contexts.
VAHC can support addressing these challenges by enabling multi-purpose, intuitive, collaborative interfaces that enhance communication, clarify roles, coordinate next steps in patient care, and integrate AI into clinical workflows, while keeping humans in the loop.

\noindent \textbf{\emph{Opportunities and Directions.}}
Most fundamentally, the challenge here is not one of interface design, but of negotiation and human-AI agency \changed{through VA support. 
To this end, VA approaches and workflows must be re-examined within AI-enabled analytical environments~\cite{vink_EuroVA_2026}}.
Future work prioritizes making responsibilities, expertise boundaries, and accountability between clinicians, patients, and AI systems explicit, inspectable, and adaptable within VAHC systems. 
This includes designing \changed{mechanisms to expose decision ownership, support bridging between human and AI actors, and enable evaluation of these dynamics.}

\paragraph{\includegraphics[height=1.1em]{figs/icons/validation.png} Model Reliability and Validation}
\changed{Reliable AI in healthcare is not a property established once before deployment; it must be continuously evaluated, monitored, and validated.} 
\changed{The workshop discussions (10/69) reveal three complementary challenges spanning the AI model lifecycle.}
\textit{Evaluation before deployment} (4/10) remains a key challenge, requiring robust benchmarks, cost-benefit considerations, and exploration of automated analysis capabilities.
A recent scoping review \changed{identifies a predominance of traditional diagnostic metrics (AUROC, sensitivity, specificity) over metrics optimized for post-deployment monitoring~\cite{andersen2024monitoring}.
Visualization research offers a direct response: Suh et al.~\cite{suh2023metrics} demonstrate that aggregate metrics alone fail to communicate a model's risks and limitations to domain experts, and that visualization-based communication supports more informed assessment of model reliability.}
\changed{\textit{Reliability after deployment} (4/10) raises the problem of} \textit{grounding}: anchoring AI outputs in validated knowledge and keeping systems up to date as conditions change.
\changed{Dataset shift, whether from population drift, technological change, or behavioral adaptation, is a failure mode clinicians are often ill-equipped to detect}, reinforcing the need for VA applications that help identify distributional change and model degradation to users~\cite{finlayson2021clinician}.
\looseness=-1
Designing reusable visual model interfaces and coordinating their evaluation across multiple patient populations remains methodologically challenging, \changed{as many tools are disease-centric and therefore limited in their generalizability across clinical contexts}.
Communication and design methodologies establishing shared language across clinical and AI experts are key to detecting and addressing dataset shift~\cite{lennerz2022unifying}.
Finally, \textit{\changed{validation under} uncertainty} (2/10) \changed{addresses a deeper problem: the standard assumption that model outputs can be verified against ground truth breaks down when expert knowledge itself is incomplete.}
\changed{Interactive VA can support expert validation in clinical settings where objective ground truth is difficult to obtain~\cite{morgenshtern2023riskfix}.}
In one VAHC 2025 example of single-cell RNA sequencing cell-type classification, authors combine multiple representations (UMAP and dot plots) with SHAP values and model weights to support reasoning under epistemic uncertainty, rather than correctness checking alone~\cite{rl2025vahc}.
Together, these examples suggest that VAHC systems can contribute to reliability not merely by displaying model outputs, but by supporting lifecycle-oriented evaluation, monitoring, and expert validation.

\noindent \textbf{\emph{Opportunities and Directions.}}
A key opportunity here lies in developing more domain-specific, context-aware model validation approaches that account for the complex, multidimensional nature of healthcare and the dynamics of these processes. 
\changed{Future work should establish lifecycle-oriented frameworks that integrate pre-deployment evaluation beyond aggregate metrics, post-deployment monitoring for dataset shift and population drift, and expert validation mechanisms for settings where ground truth is incomplete.} 

\section{Conclusion}
The VAHC 2025 workshop brought together more than 40 participants to collectively identify core challenges 
in Visual Analytics for Healthcare, with AI emerging as a recurring theme.
Through structured group discussions informed by thematic coding of the workshop's submissions, lightning talks, and open discussion prompts, we identified five AI-centered challenge clusters across all discussion themes: \textit{bias and trust}, \textit{data and infrastructure}, \textit{explainability and communication}, \textit{human-AI interaction}, and \textit{model reliability and validation}.
\changed{Across all clusters and three differently-framed discussion groups, a consistent signal emerges:}
\changed{many of} the most persistent AI challenges in healthcare VA are not primarily technical but also calibration problems, \changed{continuously striving for a balance} between trust and evidence, explainability and clinical utility, automation and human agency. 
\looseness = -1
VAHC solutions are uniquely positioned to \changed{simultaneously} bridge these tensions, operating at the intersection of data, models, and human reasoning.
We offer these challenge clusters as a synthesis of concerns and a step toward a shared research agenda, to spark discussion, research, and collaboration across the healthcare, AI, and visualization communities.
We call for the development of interoperable infrastructures, evaluation frameworks, and design practices that enable VAHC systems to support accountable, reliable, and human-centered AI in healthcare.

\section*{Acknowledgments} This research was partially funded by the {\"O}sterreichischen Wissenschaftsfonds (FWF, Austrian Science Fund) [10.55776/I6635], the Digital Society Initiative (DSI) of the University of Zurich, and
the Engineering and Physical Sciences Resource Council (EPSRC UKRI157).

\bibliographystyle{abbrv-doi-hyperref}

\bibliography{template}

\end{document}